# Positioning in groups: The roles of expertise and being in charge

Tra Huynh, Amali Priyanka Jambuge, Hien Khong, James T. Laverty, Eleanor C. Sayre
trahuynh@ksu.edu, amali@ksu.edu, hienkhong@ksu.edu, laverty@ksu.edu, esayre@ksu.edu
Department of Physics, Kansas State University

**Abstract:** Inchargeness is associated with one's authority in driving the activity in collaboration. We study how inchargeness changes within a collaborative group when its members have differing expertise. We present a case study of a group of three students working in an upper division undergraduate physics laboratory. One of them has less on-task expertise than her peers due to missing a day, which reduces her relative inchargeness across two storylines: "catching up" and "moving forward".

## Introduction

Group work is common in undergraduate physics classrooms and, if equitable, can support individual learning (Archibeque et al., 2018). We associate equity, which Esmonde (2009) defines as "fair distribution of opportunities to learn or opportunities to participate" with shared authority and power in driving group activity. We use *inchargeness* to characterize that authority to help us study group dynamics (Kustusch et al., 2018).

The inchargeness framework is rooted in Positioning Theory (Haaré & Van Langenhove, 1998). Inchargeness is intimately associated with other participants' positionings and communication acts within a storyline (Kustusch et al., 2018). An individual with high inchargeness is positioned within the group such that their acts are more likely to include successful bids to steer an activity. Paying attention to the combination of behavioral markers (who sets and limits the group activity; who asks and answers questions to direct that activity; types of discourse; whether communication acts are overlapped or sequential; and to whom the acts are done) helps us infer the group's inchargeness distribution. The group's inchargeness distribution dynamically changes over the course of interactions. However, within a determined storyline, individual inchargeness tends to shift within only a small range (Archibeque et al., 2018; Kustusch et al., 2018). While Positioning Theory is often used in concert with intersectional identity, this work focuses on inchargeness in the moment of discussion rather than how inchargeness intersects with participants' identities. Although various aspects of students and learning environment can influence inchargeness, we are interested in varying patterns of inchargeness with respect to group members' disparity in on-task expertise (relevant skills and knowledge of the task at hand).

Our video data is drawn from an upper division laboratory course at a US university where students usually spend three, three-hour lab meetings on a single experiment where they must ultimately produce individual lab reports. We studied a group of three second-year physics students – "Charlie", "Abbey", and "Will" working on the *X-ray diffraction* experiment. The lab was already set up, but the students *did* need to figure out how to operate the X-ray apparatus with different crystals and collect the data to ascertain crystals' structures. Abbey was absent for the first lab meeting when Charlie and Will explored the apparatus, made sense of the experimental principles, and collected several sets of data. During the second meeting, the group, including Abbey, seemed to perceive her lack of on-task expertise. Abbey could not immediately engage in specific on-task discussions but rather needed time to catch up with the general lab activities. Meanwhile Charlie and Will progressed in the lab and occasionally paused to explain to her what was happening. We find that the group interactions in this lab period are characteristically different from others because of a gap in their on-task expertise. Using Positioning Theory, we defined two concurrent storylines – "catching up" and "moving forward" to show how students engage in positionings and communication acts that construe their inchargeness.

## Analysis of inchargeness

The "catching up" storyline involves activities where Abbey socially and intellectually catches up with her peers regarding procedural aspects of the lab and previous lab reports. Although Abbey initiated many of the activities (requesting information, offering lab materials, etc.), she usually waited for her peers' availability to do so and received brief, dictating and expository responses. Moreover, her peers often intervened with other topics, steered the activity away, or withdrew from her topic before giving her full details (as underlined in Table 1).

The "moving forward" storyline consists of extended discussions between Charlie and Will, where they exclusively set the topics about the intellectual aspects of specific tasks needed to make progress in the lab. They sequentially asked and answered questions, built on each other's ideas to make decisions, and took actions without explicitly addressing their acts to Abbey. Charlie and Will also alternately suggested and evaluated ideas and approaches whereas Abbey gave few of such on-task suggestions or evaluations although she appeared attentive to these discussions. Occasionally, she spoke up to take bids made by her peers such as clicking the computer mouse or operating the X-ray (Table 1). Her acts throughout these discussions were not to steer the group activities away, but rather to ensure that she correctly followed up with the ones that Charlie set. Will, although he didn't speak up much in those task-assigning discourses, had built enough expertise during the previous lab meeting to allow him to intervene and successfully direct the group action as shown in Table 1.

These behavioral markers show that Abbey had significantly lower inchargeness than Charlie and Will throughout both storylines. We contend that Abbey's reduced inchargeness was the result of her low relative on-task expertise. In the "catching up" storyline, Charlie and Will were positioning as an important source of information for Abbey, from whom she actively sought responses and attention. Therefore, Charlie's and Will's acts strongly impacted the way the activity proceeded when compared to Abbey's acts as an information receiver. Throughout the "moving forward" storyline, her peers had more authority to move the lab forward, whereas Abbey played a minimum and passive contributive role.

Table 1: Excerpts from "Catching up" and "Moving forward" storylines

| "Catching up" | "Moving forward" |
| --- | --- |
| A: Do you want to do a quick review of what you've been doing?<br>W: Hm…<br>C: [to Abbey] Right now, we're calibrating.<br>W: This shoots X-ray that hits crystals [Charlie overlaps]<br>C: [to Will] Do you [want to] stand by the switch?<br>W: This is the step motor, which turns that [Charlie interrupts]<br>C: [to Will] Yep, just in case, hold on, it's still good.<br>W: Hm… [to Abbey] This piece of tape stops it if it goes pass through that kind of traffic, don't let it do that, so you shut it off before it…<br>C: [to Abbey] Yea [audio], so you have to stop.<br>A: And then what we just read the stuff off from the computer?<br>W: Yea. Hm… We already did an experiment where we look at the characteristics of X-ray off the crystal [audio].<br>[C intervenes and directs working on the experiment.] | C: [to Abbey] Alright, you can start the X-ray, and I will start acquiring the data.<br>A: So, I start now?<br>C: Yea.<br>[Abbey fails to operate the X-ray. Charlie comes over to check the X-ray.]<br>C: Hold on.<br>A: Am I doing it wrong?<br>C: No, it's me, it's just not be…<br>W: Make sure it gets on the back [audio].<br>[Charlie follows Will's instruction.]<br>C: Okay.<br>A: Ready?<br>C: Yea. |

## Conclusion

Our work investigates inchargeness with a focus on its relationship to on-task expertise to unpack the richness of group dynamics in upper division laboratories. We contend that smaller on-task expertise leads to reduced inchargeness, which may subsequently reduce opportunities for participation and learning. We observe that Abbey had much less inchargeness while Charlie and Will shared greater power in driving the group activity in "catching up" and "moving forward" storylines. In contrast, the analysis of this group in another lab where everyone attended all lab sessions and had similar on-task expertise shows more evenly distributed inchargeness. This result, even sensible, may not always be the case for all students with less expertise due to various circumstances (missing a class, joining new group, different academic background, etc). We also recognize that other factors likely influence distributions of inchargeness in collaborative groups. More work should be done to better understand these effects.

## Acknowledgments

We are greatly indebted to the instructor sharing the class observation and the copy editor *Jeremy Smith*. This work is based upon work supported by Grant ######## and a state university Physics Department.